\documentclass[english,aps,prb,twocolumn,superscriptaddress]{revtex4-1}
\usepackage{mathptmx}
\usepackage[T1]{fontenc}
\usepackage[latin9]{inputenc}
\usepackage{babel}

\usepackage{verbatim}
\usepackage{amsmath}
\usepackage{graphicx}
\usepackage{amssymb}
\usepackage[unicode=true,pdfusetitle,
 bookmarks=true,bookmarksnumbered=false,bookmarksopen=false,
 breaklinks=false,pdfborder={0 0 1},backref=false,colorlinks=false]
 {hyperref}
\usepackage{breakurl}

\makeatletter
\@ifundefined{textcolor}{}
{%
 \definecolor{BLACK}{gray}{0}
 \definecolor{WHITE}{gray}{1}
 \definecolor{RED}{rgb}{1,0,0}
 \definecolor{GREEN}{rgb}{0,1,0}
 \definecolor{BLUE}{rgb}{0,0,1}
 \definecolor{CYAN}{cmyk}{1,0,0,0}
 \definecolor{MAGENTA}{cmyk}{0,1,0,0}
 \definecolor{YELLOW}{cmyk}{0,0,1,0}
 }

\PassOptionsToPackage{normalem}{ulem}
\usepackage{ulem}
\let\cite@rig\cite
\newcommand{\b@xcite}[2][\%]{\def\def@pt{\%}\def\pas@pt{#1}
  \mbox{\ifx\def@pt\pas@pt\cite@rig{#2}\else\cite@rig[#1]{#2}\fi}}
\renewcommand{\underbar}[1]{{\let\cite\b@xcite\uline{#1}}}

\makeatother

\begin{document}

\title{From surface to volume plasmons in hyperbolic metamaterials: General
existence conditions for bulk high-$k$ waves in metal-dielectric
and graphene-dielectric multilayers}

\author{Sergei V. Zhukovsky}

\email{sezh@fotonik.dtu.dk}

\affiliation{DTU Fotonik, Department of Photonics Engineering, Technical University
of Denmark, {\O}rsteds Plads 343, DK-2800 Kgs. Lyngby, Denmark}

\author{Andrei Andryieuski}

\affiliation{DTU Fotonik, Department of Photonics Engineering, Technical University
of Denmark, {\O}rsteds Plads 343, DK-2800 Kgs. Lyngby, Denmark}

\author{J. E. Sipe}

\affiliation{Department of Physics and Institute for Optical Sciences, University
of Toronto,\\
60 St.~George Street, Toronto, Ontario M5S 1A7, Canada}

\author{Andrei V. Lavrinenko}

\affiliation{DTU Fotonik, Department of Photonics Engineering, Technical University
of Denmark, {\O}rsteds Plads 343, DK-2800 Kgs. Lyngby, Denmark}
\begin{abstract}
We theoretically investigate general existence conditions for broadband
bulk large-wavevector (high-$k$) propagating waves (such as volume
plasmon polaritons in hyperbolic metamaterials) in subwavelength periodic
multilayer structures. Describing the elementary excitation in the
unit cell of the structure by a generalized resonance pole of a reflection
coefficient, and using Bloch's theorem, we derive analytical expressions
for the band of large-wavevector propagating solutions. We apply our
formalism to determine the high-$k$ band existence in two important
cases: the well-known metal-dielectric, and recently introduced graphene-dielectric
stacks. We confirm that short-range surface plasmons in thin metal
layers can give rise to hyperbolic metamaterial properties, and demonstrate
that long-range surface plasmons cannot. We also show that graphene-dielectric
multilayers tend to support high-$k$ waves and explore the range
of parameteres for which this is possible, confirming the prospects
of using graphene for materials with hyperbolic dispersion. The approach
is applicable to a large variety of structures, such as continuous
or structured microwave, terahertz (THz) and optical metamaterials%
{}.
\end{abstract}


\maketitle

\section{Introduction}

Hyperbolic metamaterials (HMMs) are composite media that consist of
subwavelength structures assembled so that an extreme anisotropy results
on the macroscopic scale, with metallic behavior arising for one polarization
of light and dielectric behavior for the other. In other words, their
permittivity tensor $\varepsilon=\text{diag}(\epsilon_{x},\epsilon_{y},\epsilon_{z})$
has eigenvalues of different signs (e.g., $\epsilon_{x}=\epsilon_{y}<0$
and $\epsilon_{z}>0$ in the uniaxial geometry). Such anisotropy results
in the dispersion relation in such a medium \begin{equation}
\frac{\omega^{2}}{c^{2}}=\frac{k_{x}^{2}+k_{y}^{2}}{\epsilon_{z}}+\frac{k_{z}^{2}}{\epsilon_{x,y}}\label{eq:disp}\end{equation}
that is hyperbolic rather than elliptical (Fig.~1a), hence the name
of HMMs. A hyperboloidal isofrequency surface is much more extended
in the wave vector space than an ellipsoidal one -- indeed, theoretically
infinite in the idealization that Eq.~\eqref{eq:disp} holds for
all $k_{x,y,z}$ -- so an HMM supports propagating solutions with
very large wave vectors ($k^{2}\gg\epsilon_{x,y,z}\omega^{2}/c^{2}$).
These waves, called high-$k$ waves for short \cite{JacobAPB10},
would be extremely evanescent in any natural isotropic or weakly birefringent
medium but become propagating in HMMs. The existence of high-$k$
waves brings about a rich variety of new physics, both related to
the waves themselves (as highly confined information carriers for
subwavelength imaging \cite{Hyperlens06}) and associated with a tremendous
increase in the photonic density of states (PDOS) in HMMs, resulting
in strong modification of all light-matter interaction phenomena that
depend on it, such as spontaneous emission \cite{NarimanovAPL12}.

What truly sparked the explosive scientific interest during the past
few years was the discovery that HMM functionality can be exhibited
in a non-resonant, broadband manner by structures with very simple
geometry, such as nanorod arrays \cite{NoginovAPL09,NoginovOL10}
and metal-dielectric multilayers \cite{JacobAPB10,NarimanovAPL12}.
An anomalous increase of the decay rate of nearby emitting centers
(a broadband Purcell effect) was demonstrated experimentally \cite{NoginovOL10,JacobAPB10},
along with the direct measurement of radiation enhancement \cite{RecentKimOE12}.
Many applications of HMMs have ben suggested, such as far-field subwavelength
imaging or ``hyperlensing'' \cite{Hyperlens06} and highly absorptive
surfaces that benefit (rather than suffer) from increased roughness
\cite{NarimanovBlackOE13}. More fundamental and more intriguing uses
for HMMs have also been envisaged, exploiting mathematical similarities
between sign changes in the dispersion relation \eqref{eq:disp} and
metric signature transitions in cosmological equations \cite{SmolPRL10,SmolJOSA11}.
Many more areas of research are being explored, as can be seen in
the recent reviews \cite{ReviewJacob,ReviewKildishev} and references
therein.

\begin{figure}
\includegraphics[width=1\columnwidth]{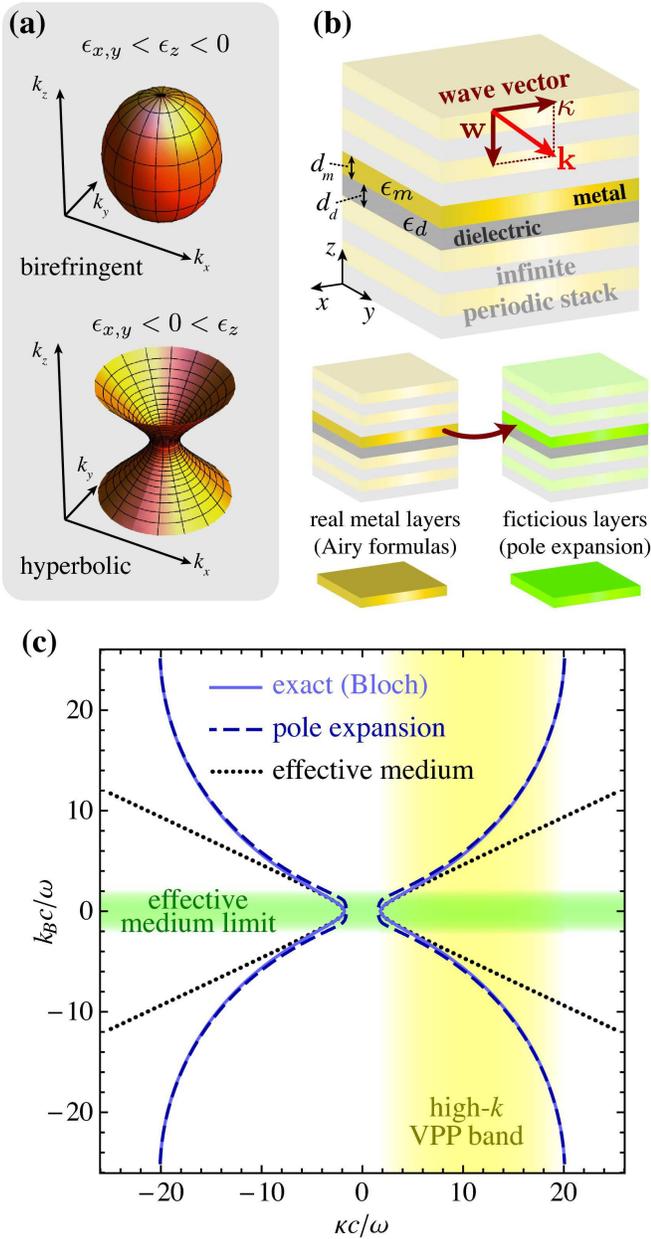}

\caption{(Color online) Theoretical background on hyperbolic metamaterials
(HMMs). (a) Isofrequency surfaces in the dispersion relation {[}Eq.~\eqref{eq:disp}{]}
for conventional anisotropic medium ($\epsilon_{x,y,z}>0$) and HMM
($\epsilon_{x,y}<0$ and $\epsilon_{z}>0$). (b) An infinite periodic
multilayer HMM with geometric notations and wave vector decomposition
used in the paper, along with the illustration of the replacement
of real metal layers with reflection and transmission coefficients
given by Eq.~\eqref{eq:airy} with fictitious layers featuring just
a pole-like elementary excitation with reflection and transmission
coefficients given by Eq.~\eqref{eq:pole}. (c) Comparison between
the actual multilayer dispersion relation {[}Eq.~\eqref{eq:bloch}{]},
the dispersion relation derived from the pole expansion {[}Eq.~\eqref{eq:poledisp}{]},
and the effective-medium dispersion relation {[}Eq.~\eqref{eq:disp}
with Eq.~\eqref{eq:homo}{]} for the structure with layer thicknesses
$d_{m}=$2.3~nm, $d_{d}=$11.4~nm and permittivities $\epsilon_{m}=-17.2$,
$\epsilon_{d}=2.59$ (relevant for Au/$\mathrm{Al}_{2}\mathrm{O}_{3}$
structures for $\lambda=$715 nm\cite{JacobAPB10}). The yellow shaded
area shows the band of propagating high-$k$ VPPs. \label{FIG:Schematic}}

\end{figure}

Even though the effective permittivity representation of HMMs has
proved very successful in predicting and explaining their exotic physics,
it is the high-$k$ waves that govern the functioning of any HMM on
a microscopic level. Hence it is these waves that eventually determine
the extents and limits of applicability of a particular HMM with respect
to any of the effects described above. Thus, it is crucial to understand
the physical nature of these waves. In metal-dielectric structures,
the conventional wisdom is that the nature is plasmonic, so various
groups have chosen different terms for them: \emph{multilayer plasmons}
\cite{Schilling06}, \emph{Bloch plasmon polaritons} \cite{Avrutsky07},
or \emph{volume plasmon polaritons} (VPPs) \cite{ReviewKildishev,LasPhotRev13}.
In HMMs with a multilayer geometry, VPPs should arise from coupling
of surface plasmon polaritons (SPPs) at layer interfaces \cite{plasFeng05,plasPendry06,plasOrenstein}.
In our recent work \cite{ourFocusOE13}, we showed explicitly that
VPPs originate from coupling of short-range SPPs (SRSPPs) in individual
metal layers by keeping only the SRSPP response in these layers via
a pole expansion. It is noteworthy that an SRSPP exists for just one
value of the wave vector, whereas the resulting VPPs exist in the
entire range of them, spanning the isofrequency surface in Fig.~\ref{FIG:Schematic}c.%

Two interesting observations were made alongside this proof. First,
it appeared that there is a stark contrast between the two characteristic
excitation in the metal layer: the short-range SPP capable of giving
rise to HMM behavior, and the long-range SPP (LRSPP) that do not have
such a capability. Second, as also mentioned in other accounts \cite{YuryNLmultiple11},
VPPs were shown to exist outside of the HMM regime, albeit in a somewhat
narrower band in the wave vector space. The general principle, namely,
{}``lower-dimensional elementary unit cell excitations coupling form
a higher-dimensional excitation in a periodic arrangement of such
cells'' is undoubtedly behind the formation of VPPs in multilayer
HMMs. Nevertheless, it still remains to be determined what conditions
these elementary excitations must satisfy to form a high-$k$ band
spanning a broad range of $k$, such as happens in VPPs. A general
understanding would be very useful in determining the applicability
range for new types of HMMs, such as, for example, graphene-based
multilayers introduced in recent works \cite{grapheneAndrei,grapheneBelov,grapheneCapolino,grapheneStrangi}. 

In this paper,\textbf{ }we theoretically investigate general existence
conditions for broadband bulk high-$k$ propagating waves (such as
VPPs in HMMs) in arbitrary periodic multilayers structures. We treat
the elementary excitation in the unit cell of such a structure as
a generalized resonance defined by a pole-like response in its Fresnel
reflection and transmission coefficients. Then, using Bloch's theorem,
we derive analytical expressions for the band of high-$k$ propagating
solutions that can originate from this elementary excitation by hybridization
in the periodic structure. Using these analytical expressions, we
show that SRSPPs in thin metal layers can -- and commonly do -- give
rise to HMM-like properties in subwavelength metal-dielectric multilayers;
on the other hand, LRSPPs only form a very narrow plasmonic band near
the light line of the dielectric and do not produce a high-$k$ band.
Furthermore, we apply the formalism to graphene-dielectric metamaterials
in the THz range. We show that TM-polarized plasmons in individual
graphene sheets also hybridize to form VPPs with HMM-like properties
in the frequency range where the imaginary part of the graphene conductivity
significantly exceeds its real part. On the other hand, \emph{transverse}
or TE-polarized graphene plasmons \cite{TEgraphene2007,TEgrapheneOE}
behave like LRSPPs in metal-dielectric multilayers, not giving rise
to HMM-like behavior. 

The present results are primarily valuable from the theoretical point
of view, providing a general understanding of how high-$k$ band of
bulk propagating waves originate from fixed-$k$ surface excitations
in individual layers of a multilayer system. In addition, our results
have practical applications, allowing for very efficient estimation
of VPP dispersion and HMM properties in existing HMMs (metal-dielectric
and graphene-dielectric multilayers), which is useful in the design
of HMM-based devices such as hyperlenses with improved performance.
Moreover, the present results provide a means to determine whether
any localized excitation (electromagnetic or otherwise) is likely
to give rise to HMM-like or bulk plasmon-like behavior when assembled
into a periodic system. Examples may include optical waveguide arrays,
multilayers supporting Bloch surface waves or spoof surface plasmons,
periodic layers of two-dimensional electron gas (e.g. multiple quantum-well
semiconductor heterostructures) or even acoustic multilayers. 

This paper is organized as follows.\textbf{ }Section \ref{sec:REVIEW}
briefly reviews the basic theoretical background on multilayer HMMs,
including the dispersion relation for the VPPs and its representation
using the pole expansion formalism \cite{ourFocusOE13}. In Section
\ref{sec:CENTRAL} we analyze this representation and derive the existence
conditions for broad VPP band formation from arbitrary resonant elementary
excitation in the metamaterial's unit cell. In Section \ref{sec:APPLICATIONS}
we apply our findings to several cases of pole expansion, including
multilayers made of metal and graphene; specifically, we show that
short-range SPPs do give rise to HMM behavior while long-range SPPs
do not. We also analyse the applicability of graphene for high-$k$
HMMs. Finally, in Section \ref{sec:CONCLUSIONS} we summarize the
results.

\section{Volume plasmon polaritons \protect \\
in multilayer hyperbolic metamaterials\label{sec:REVIEW}}

We begin by recalling that in an infinite, periodic metal-dielectric
structure (Fig.~1b) where losses are neglected, and so metal layers
with permittivity $\epsilon_{m}<0$ and thickness $d_{m}$ alternate
with dielectric layers with permittivity $\epsilon_{d}>0$ and thickness
$d_{d}$, the dispersion relation of propagating waves can be obtained
via Bloch's theorem using the standard transfer matrix approach \cite{BookYarivYeh}.
From the transfer matrix for one period of the structure,\begin{equation}
M_{1}=\frac{1}{T_{m}}\left[\begin{array}{cc}
T_{m}^{2}-R_{m}^{2} & R_{m}\\
-R_{m} & 1\end{array}\right]\left[\begin{array}{cc}
e^{iw_{d}d_{d}} & 0\\
0 & e^{-iw_{d}d_{d}}\end{array}\right],\label{eq:transmat}\end{equation}
where the reflection and transmission coefficients of a metal layer
$R_{m}$ and $T_{m}$ are given by the Airy formulas,\begin{equation}
R_{m}=r_{dm}+\frac{t_{dm}r_{md}t_{md}e^{2iw_{m}d_{m}}}{1-r_{md}^{2}e^{2iw_{m}d_{m}}},\;\; T_{m}=\frac{t_{dm}t_{md}e^{iw_{m}d_{m}}}{1-r_{md}^{2}e^{2iw_{m}d_{m}}},\label{eq:airy}\end{equation}
the interface reflection and transmission coefficients given by the
Fresnel formulas,\begin{equation}
\begin{gathered}r_{md}=\frac{w_{m}\epsilon_{d}-w_{d}\epsilon_{m}}{w_{m}\epsilon_{d}+w_{d}\epsilon_{m}},\; r_{dm}=\frac{w_{d}\epsilon_{m}-w_{m}\epsilon_{d}}{w_{d}\epsilon_{m}+w_{m}\epsilon_{d}},\\
t_{md}=\frac{2w_{m}\sqrt{\epsilon_{m}\epsilon_{d}}}{w_{m}\epsilon_{d}+w_{d}\epsilon_{m}},\; t_{dm}=\frac{2w_{d}\sqrt{\epsilon_{d}\epsilon_{m}}}{w_{d}\epsilon_{m}+w_{m}\epsilon_{d}},\end{gathered}
\label{eq:fresnel}\end{equation}
and \begin{equation}
w_{m}=\sqrt{\epsilon_{m}\omega^{2}/c^{2}-\kappa^{2}},\quad w_{d}=\sqrt{\epsilon_{d}\omega^{2}/c^{2}-\kappa^{2}},\label{eq:w}\end{equation}
expressing the relation between the tangential ($\kappa$) and the
normal ($w$) component of the wave vector in each layer (we choose
the square root of complex $w_{j}$ so that $\mathrm{Im\,}w_{j}\geq0$;
if $\mathrm{Im\,}w_{j}=0$ we take $\mathrm{Re\,}w_{j}\geq0$), Bloch's
theorem yields \cite{plasFeng05,plasOrenstein,KildishevOE11}\begin{equation}
\begin{gathered}\frac{\mathrm{Tr}\, M_{1}}{2}=\cos\left[k_{B}(d_{m}+d_{d})\right]=\cos(w_{m}d_{m})\cos(w_{d}d_{d})\\
-\frac{1}{2}\left(\frac{\epsilon_{m}w_{d}}{\epsilon_{d}w_{m}}+\frac{\epsilon_{d}w_{m}}{\epsilon_{m}w_{d}}\right)\sin(w_{m}d_{m})\sin(w_{d}d_{d}).\end{gathered}
\label{eq:bloch}\end{equation}

This expression describes a propagating Bloch wave with tangential
component $\kappa$ and normal component $k_{B}$. To relate it to
the hyperbolic dispersion relation \eqref{eq:disp}, one can Taylor
expand it around the points where $\cos[k_{B}(d_{m}+d_{d})]=1$. Provided
the layers are thin enough ($d_{m},d_{d}\ll\lambda$) so that $w_{j}d_{j}\ll1$,
Eq.~\eqref{eq:bloch} reduces to $\omega^{2}/c^{2}=k_{B}^{2}/\epsilon_{x,y}+\kappa^{2}/\epsilon_{z}$
\cite{Schilling06}, similar to Eq.~\eqref{eq:disp} where $k_{B}=k_{z}$,
$\kappa^{2}=k_{x}^{2}+k_{y}^{2}$, and \begin{equation}
\epsilon_{x}=\epsilon_{y}=\frac{d_{m}\epsilon_{m}+d_{d}\epsilon_{d}}{d_{m}+d_{d}},\quad\epsilon_{z}^{-1}=\frac{d_{m}\epsilon_{m}^{-1}+d_{d}\epsilon_{d}^{-1}}{d_{m}+d_{d}},\label{eq:homo}\end{equation}
which results in $\epsilon_{x}=\epsilon_{y}<0$ and $\epsilon_{z}>0$
in a broad range of parameters. 

As seen in Fig.~\ref{FIG:Schematic}c, the Bloch waves in HMMs approach
a hyperbolic dispersion relation for smaller $\kappa$ but deviate
from it as $\kappa$ increases. The upper limit $\kappa_{\text{max}}\propto1/(d_{m}+d_{d})$
\cite{ourPRA12} is a cut-off imposed by the finite thickness of constituent
layers, fundamentally limiting the applicability of the effective
medium approximation. This cut-off, essentially resulting from violation
of the subwavelength condition for waves with very large $\kappa$,
is the primary limiting factor for the overall PDOS increase in multilayer
HMMs. Another limitation is related to the presence of losses \cite{ReviewKildishev},
which would also eventually render the Bloch waves fully evanescent.
Yet anther, more fundamental limitation is associated with the non-local
effects in the response of the electron gas in metal \cite{AsgerPRB}. 

Finally, to relate the Bloch waves given by Eq.~\eqref{eq:bloch}
to VPPs, one can replace the metal layers with a hypothetical structure
whose reflection and transmission coefficients only contain one resonant
guided-wave excitation, i.e., are of the form of a simple pole \cite{ourFocusOE13}\begin{equation}
T_{m}=\frac{\tau}{\kappa-\kappa_{p}},\quad R_{m}=\frac{-\tau}{\kappa-\kappa_{p}}-\frac{\tau}{\kappa_{p}},\label{eq:pole}\end{equation}
where the location of the pole $\kappa_{p}$ and the pole strength
$\tau$ depend on the exact nature of the excitation. Practically,
for the case of metal-dielectric multilayers one can determine $\tau$
and $\kappa_{p}$ by comparing the generalized form of reflection
and transmission coefficients Eq.~\eqref{eq:pole} to the Airy formulas\,\eqref{eq:airy}.
The second term in the expansion for $R_{m}$ is to ensure that $R_{m}=0$
for $\kappa\to0$, as normally incident light should be incapable
of exciting any guided wave in a planar layer due to momentum conservation.

Calculating the values of $\kappa_{p}$ and $\tau$ for the SRSPP
in a thin metal layer, and substituting Eq.~\eqref{eq:pole} rather
than Eq.~\eqref{eq:airy} into Eq.~\eqref{eq:transmat}, one obtains
a modified form of the dispersion relation \cite{ourFocusOE13},\begin{equation}
\cos\left[k_{B}(d_{m}+d_{d})\right]\approx\left[1-\frac{\kappa-\kappa_{p}}{2\tau}\right]e^{iw_{d}d_{d}}+\frac{\kappa-\kappa_{p}}{2\tau}e^{-iw_{d}d_{d}},\label{eq:poledisp}\end{equation}
which is seen to correspond to the exact dispersion relation of the
multilayer very closely (Fig.~\ref{FIG:Schematic}c), correctly describing
high-$k$ waves in such a multilayer. Hence, it can be concluded that
these waves originate from hybridization of SRSPPs in the metal layers,
and are indeed VPPs. We note that these VPPs exist in a very wide
range of $\kappa$ while the original SPPs only exist at a single
$\kappa=\kappa_{p}$. It is somewhat surprising that VPP formation
is specific to SRSPPs; other pole-like excitations present in a thin
metal layer, such as the LRSPP, was not found to affect the VPP band
in any significant way. Another interesting observation was that VPPs
were found to exist \emph{outside} of the HMM range \cite{ourFocusOE13},
where earlier works indeed predicted two branches of propagating waves,
one with positive and one with negative refraction \cite{YuryNLmultiple11}.
However, the range of the VPP band in this regime was found to be
significantly narrower in $\kappa$, and Eq.~\eqref{eq:disp} was
no longer applicable. 

All these facts taken together mean that it is necessary to understand
whether and when a pole-like excitation of the type of Eq.~\eqref{eq:pole}
leads to the formation of a VPP band which would be sufficiently broadband
to lead to an HMM response. This understanding, in the form of generalized
existence conditions for the VPP band, is especially important for
predicting whether an HMM regime is possible with new types of plasmonic
multilayer structures, such as those based on graphene, and what prevents
other kinds of structures, such as high-index dielectric waveguide
arrays, from giving rise to HMM properties.

\section{Formation of large-wavevector band\label{sec:CENTRAL}}

We begin our investigation by using Eq.~\eqref{eq:pole} and rederiving
the dispersion relation in the general form, \begin{equation}
\begin{gathered}\cos\left[k_{B}(d_{m}+d_{d})\right]=\frac{\kappa-\kappa_{p}}{2\tau}e^{\frac{2\pi d}{\lambda}\sqrt{\kappa^{2}-\epsilon_{d}}}\\
-\left[\frac{\tau}{\kappa_{p}}+\left(\frac{\tau}{\kappa_{p}}\right)^{2}\frac{\kappa-\kappa_{p}}{2\tau}\right]e^{-\frac{2\pi d}{\lambda}\sqrt{\kappa^{2}-\epsilon_{d}}}\equiv F(\kappa),\end{gathered}
\label{eq:newdisp}\end{equation}
where $\tau$, $\kappa$ and $\kappa_{p}$ are now dimensionless (normalized
by $\omega/c=2\pi/\lambda$). 

The existence condition for propagating waves will then be $F(\kappa)\in[-1;1]$,
and in order to better analyze it, we introduce several dimensionless
quantities: \begin{equation}
\xi\equiv\frac{\tau}{\kappa_{p}},\;\eta\equiv\frac{2\pi d}{\lambda},\;\chi\equiv\frac{\sqrt{\epsilon_{d}}}{\kappa_{p}},\;\text{and}\;\beta\equiv\frac{\kappa}{\kappa_{p}}.\label{eq:dimensionless}\end{equation}
We see that $\eta$ is the measure of how {}``subwavelength'' the
spacer dielectric layers appear to be with respect to the vacuum wavelength
of the incident light (so normally $\eta\ll1$); $\chi$ indicates
the position of the dielectric cut-off (point on the light line for
a given frequency) normalized to the position of the pole (again,
$\chi\ll1$ and can be neglected unless $\kappa_{p}$ is very close
to the light line, e.g., for LRSPPs); $\xi$ characterizes the pole
strength (and nothing will be assumed about it; note that Eq.~\eqref{eq:poledisp}
results from Eq.~\eqref{eq:newdisp} for $\xi\to-1$); and $\beta$
is the tangential component of the wavevector normalized to the position
of the pole. Using these quantities, we can rewrite Eq.~\eqref{eq:newdisp}
in a more symmetric way,\begin{equation}
\tfrac{1}{2}(\beta-1)A(\beta)-\tfrac{1}{2}(\beta+1)A^{-1}(\beta)\equiv F(\beta)\in[-1;1],\label{eq:central}\end{equation}
where \begin{equation}
A=\xi^{-1}\exp\left(\eta\kappa_{p}\sqrt{\beta^{2}-\chi^{2}}\right).\label{eq:A}\end{equation}

We can now analyze the limiting cases of Eq.~\eqref{eq:central}.
For the combination of parameters such that $A\gg1$ (achieved for
significantly large $\kappa_{p}$ and/or very small $\xi$), one can
neglect the term with $A^{-1}$ and see that $F(\beta)=\pm1$ can
be solved analytically to yield \begin{equation}
\beta=1+\left(\eta\kappa_{p}\right)^{-1}W\left(\pm2\xi\eta\kappa_{p}e^{-\eta\kappa_{p}}\right),\label{eq:ass1_beta}\end{equation}
where $W(z)$ is the Lambert \emph{W} function \cite{LambertW} defined
as the solution of $W(z)e^{W(z)}=z$. If its argument is small, we
can use the first-order approximation $W(z)\approx z$ to yield that
VPPs exist if \begin{equation}
\beta\in\left[1-2\xi e^{-\eta\kappa_{p}};\,1+2\xi e^{-\eta\kappa_{p}}\right].\label{eq:ass1_beta1}\end{equation}
 So, there should be a VPP band around $\beta=1$ (i.e., $\kappa=\kappa_{p}$),
which is usually narrow unless $\eta\kappa_{p}$ is small. This explains
that the VPP band should widen as the structure becomes more subwavelength
(smaller $\eta$), but does not explain how it can fill a very broad
range of $\kappa$ since that would require violating the assumptions
leading to Eq.~\eqref{eq:ass1_beta1}. Additionally, we see that
$\beta=1$ represents the limiting behavior of Eq.~\eqref{eq:central}
for very small $\xi$.

The other limit can be obtained by noticing that for the case of $A=A^{-1}=1$,
we have $F(\beta)=-1$ regardless of any other parameters. Changing
to the logarithmic scale with respect to $\xi$ by defining $x\equiv\ln|\xi|$,
we see that the dependence \begin{equation}
x=\eta\kappa_{p}\beta\;\text{where }e^{x}=|\xi|\label{eq:ass2_x}\end{equation}
is likely to describe the edge of another VPP band. Indeed, if the
argument of the exponent in Eq.~\eqref{eq:A} is small enough so
that $A^{\pm1}\approx1\pm(\eta\kappa_{p}\beta-x)$, then it follows
that $F(\beta)=0$ for $x=\eta\kappa_{p}\beta-1/\beta$ and $F(\beta)=1$
for $x=\eta\kappa_{p}\beta-2/\beta$. Hence Eq.~\eqref{eq:ass2_x}
describes the second limiting case for the existence of the VPP band,
which is narrow for large $\xi$ (and $\beta$) and becomes wider
as $\xi$ decreases.

\begin{figure}
\includegraphics[width=1\columnwidth]{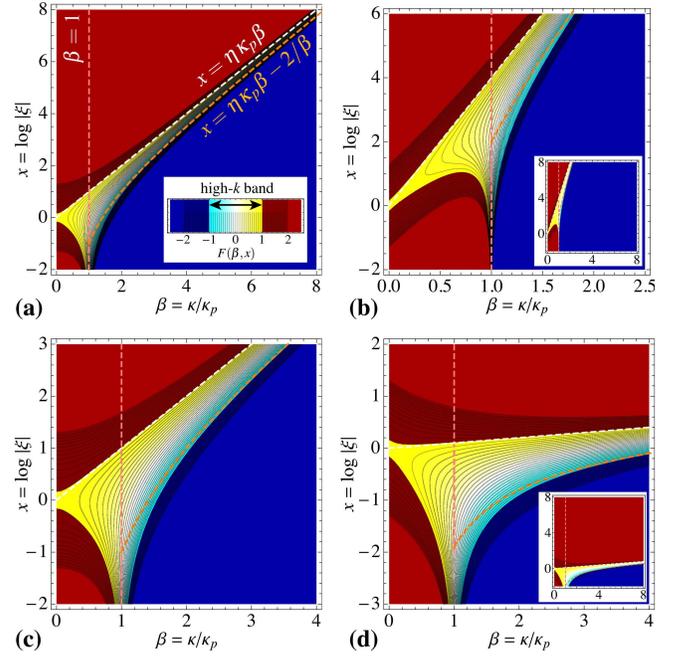}\caption{(Color online) Behavior of Eq.~\eqref{eq:central} in different regimes.
(a) Illustration of the limiting behavior of $F(\beta,x=\log|\xi|)$
for large $\beta$ and $x$; the dashed lines show the asymptotes
given by Eqs.~\eqref{eq:ass1_beta1}--\eqref{eq:ass2_x}. (b--d)
An enlarged view of $F(\beta,x)$ around the asymptote intersection
point $(\beta=1,x=\eta\kappa_{p})$ for $\eta\kappa_{p}$ equal to
(b) 4, (c) 1, and (d) 0.1. The insets show the plots in (b) and (d)
in the same scale as (a) for comparison.\label{FIG:Central}}
\end{figure}

This limiting case analysis can be used to easily predict the existence
of the VPP band in the $(\beta,x)$ space, as seen in Fig.~\ref{FIG:Central}a.
We see that the VPP band follows the line $\beta=1$ until $x=0$,
after which it slants to follow the line $\beta=x/(\eta\kappa_{p})$.
The VPP bands corresponding to the two asymptotes overlap near the
intersection point ($\beta=1$, $x=\eta\kappa_{p}$), where the equation
ceases to be analytically solvable and the VPP band has a more complicated
shape (Fig.~\ref{FIG:Central}b--d). It was also established that
non-zero $\zeta$ influences this behavior only weakly, suppressing
any solutions of Eq.~\eqref{eq:central} below the light line ($\beta\leq\chi$)
and slightly modifying the limiting behavior near it.

We can thus identify two distinct characteristic cases when the VPP
band is sufficiently broad. First, it can be seen that for the both
limiting cases, the band widens as $\xi$ approaches unity (Fig.~\ref{FIG:Central}).
Hence the region $0.1\leq\xi\leq10$ corresponds to the case when
bulk propagating solutions are supported for the widest range of $\kappa$.
Second, for the special case $\eta\kappa_{p}\ll1$, when the line
corresponding to Eq.~\eqref{eq:ass2_x} is nearly horizontal, there
is a narrow range of $x\in[-2/\beta;0]$ when the VPP band width spans
from below unity all the way to $\sqrt{2/(\eta\kappa_{p})}$, approaching
very large values for $x\to-0$ for very deeply subwavelength structures.

\section{Examples\label{sec:APPLICATIONS}}

\subsection{SRSPP and LRSPP in metal-dielectric stacks}

The most straightforward way to test the proposed criteria is to apply
the conditions to the well-studied HMM produced from metal-dielectric
multilayers, using SPPs in the metal layers as the elementary excitations
in Eq.~\eqref{eq:pole}. It is known that a metal layer supports
two types of such plasmons depending on whether the individual plasmons
at the layer interfaces are coupled symmetrically or antisymmetrically
with respect to the dominant field component $E_{z}$. Both these
modes can be obtained from the equation \begin{equation}
1-r_{md}^{2}\exp(2iw_{m}d_{m})=0.\label{eq:polecond}\end{equation}

The primary difference between them is the behavior of their propagation
constant $\kappa_{p}$ as the metallic layer thickness $d_{m}$ approaches
zero. The symmetrically coupled SPP has its wave vector approach the
light cone ($\kappa_{p}\to\sqrt{\epsilon_{d}}$), and if the metal
is lossy, the losses decrease as the wave becomes increasingly less
confined to the layer. The asymmetrically coupled SPP has its wave
vector approach infinity ($\kappa_{p}\to\infty$), and the wave becomes
increasingly more confined to the metal layer, so the losses increase.
For the latter reason, these two SPPs are traditionally denoted long-range
and short-range, or LRSPP and SRSPP, respectively. 

\begin{figure}
\includegraphics[width=1\columnwidth]{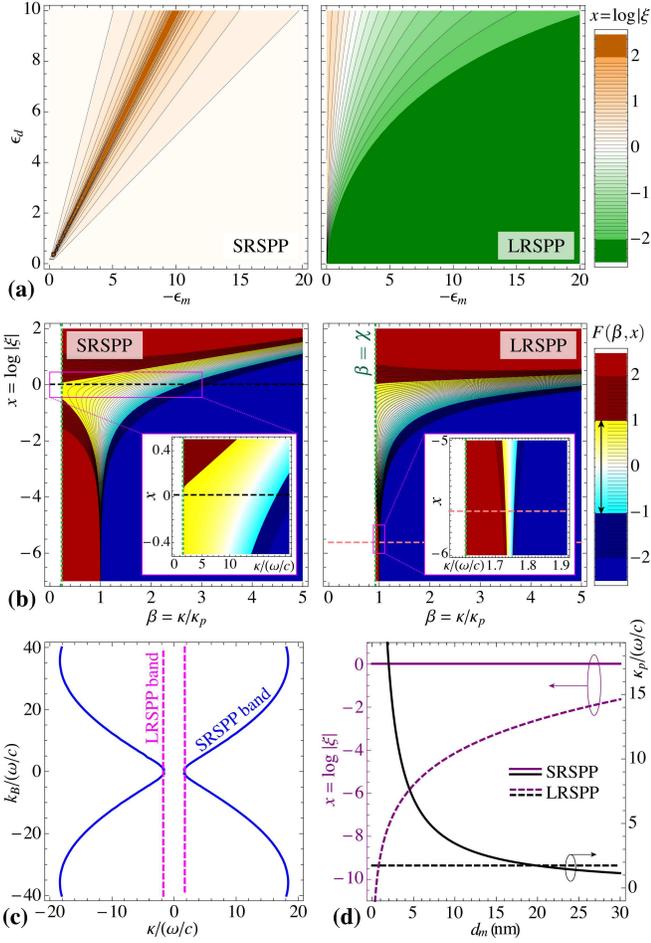}\caption{(Color online) Formation of the VPP band in metal-dielectric HMMs
from SRSPPs and LRSPPs. (a): Dependence of $x$ on the permittivities
of metal and dielectric in a thin metal layer ($d_{m}=5$~nm) for
SRSPPs and LRSPPs. (b) Example dependencies of $F(\beta,x)$ for SRSPP
and LRSPP in a structure with $d_{m}=d_{d}=5$~nm and material parameters
as in Fig.~\ref{FIG:Schematic}c. The insets show the enlarged view
in the scale of $\kappa_{p}/(\omega/c)$. (c--d) Comparison between
SRSPP and LRSPP case for (c) the dispersion relation of the VPP band
and (d) the dependence of $\kappa_{p}$ and $x$ on $d_{m}$.\label{FIG:Metal}}
\end{figure}

It has already been proved \cite{ourFocusOE13} that SRSPPs can and
do give rise to the VPP band in metal-dielectric HMMs, and we begin
by reproducing this result with the proposed criteria. In the appropriate
limit of sufficiently thin metal layer, the expressions for the pole
expansion parameters for Eq.~\eqref{eq:pole} can be obtained from
Eq.~\eqref{eq:polecond} as, in dimensionless units,\begin{equation}
\kappa_{p}=\frac{\log|r|}{2\pi d_{m}/\lambda},\;\tau=\frac{r^{-1}-r}{2(2\pi d_{m}/\lambda)},\; r=\lim_{\kappa\to\infty}r_{md}=\frac{\epsilon_{m}-\epsilon_{d}}{\epsilon_{m}+\epsilon_{d}}.\label{eq:SR_coefs}\end{equation}
We note at once that $\xi$ does not depend on $d_{m}$, making the
analysis particularly easy:\begin{equation}
\xi=\frac{r^{-1}-r}{2\log|r|}=\frac{2f}{(f^{2}-1)\log\left|\frac{1+f}{1-f}\right|},\label{eq:SR_xi}\end{equation}
where $f=-\epsilon_{d}/\epsilon_{m}$. We can see that $\xi\gtrsim1$
unless $f\to1$ (Fig.~\ref{FIG:Metal}a). So, it can be concluded
that broadband VPPs are commonly formed by hybridization of SRSPPs,
as confirmed by the example in Fig.~\ref{FIG:Metal}b. The only exception
is when $\epsilon_{m}+\epsilon_{d}\approx0$ leading to $|\xi|\gg1$,
so the VPP band becomes increasingly more narrow-band and moves towards
larger $\kappa$ (see the inset in Fig.~\ref{FIG:Central}a). Note
that this corresponds to an epsilon-near-zero (ENZ) regime rather
than an HMM regime according to Eqs.~\eqref{eq:homo}, and the narrowing
and shifting of the VPP band near the ENZ points is consistent with
our earlier observation \cite{ourFocusOE13}. The VPP band shift remains
small since the slope parameter, $\eta\kappa_{p}=(d_{d}/d_{m})\log|r|$,
becomes very large in the ENZ case. 

On the other hand, the LRSPP is obtained from Eq.~\eqref{eq:polecond}
by considering the other limit ($\kappa_{p}\to\sqrt{\epsilon_{d}}$).
The resulting expressions are \begin{equation}
\begin{gathered}\kappa_{p}=\sqrt{\epsilon_{d}+(\epsilon_{d}-\epsilon_{m})\frac{\epsilon_{d}^{2}}{\epsilon_{m}^{2}}\left(\frac{1-\delta}{1+\delta}\right)^{2}},\\
\tau=\frac{\epsilon_{d}-\epsilon_{m}}{2\kappa_{p}}\frac{\epsilon_{d}^{2}}{\epsilon_{m}^{2}}\frac{(\delta-1)^{2}}{\delta(\delta+1)},\;\delta=e^{-d_{m}\sqrt{(\epsilon_{d}-\epsilon_{m})}},\end{gathered}
\label{eq:LR_coefs}\end{equation}
which gives \begin{equation}
\xi=\frac{(\epsilon_{d}-\epsilon_{m})\epsilon_{d}^{2}/\epsilon_{m}^{2}}{\epsilon_{d}+(\epsilon_{d}-\epsilon_{m})\frac{\epsilon_{d}^{2}}{\epsilon_{m}^{2}}\left(\frac{1-\delta}{1+\delta}\right)^{2}}\frac{(\delta-1)^{2}}{\delta(\delta+1)},\label{eq:LR_xi}\end{equation}
and it can be seen that $\xi\to0$ as $d_{m}\to0$ and the structure
becomes increasingly more subwavelength. This means that LRSPPs hybridize
to form but a very narrow band around $\kappa_{p}$ according to Eq.~\eqref{eq:ass1_beta1}
(Fig.~\ref{FIG:Central}a), and thus do not contribute to the VPP
band. This is seen in Fig.~\ref{FIG:Metal}a, and further demonstrated
by comparing the location of the area given by $-1<F(\beta,x)<1$
in the $(\beta,x)$ coordinates for the characteristic metal-dielectric
multilayers (Fig.~\ref{FIG:Metal}b) and the dispersion of VPPs in
the bands (Fig.~\ref{FIG:Metal}c) for LRSPP vs.~SRSPP cases. This
result can also be explained by noting that an LRSPP in a metallic
layer bears more and more resemblance to a plane wave in the surrounding
medium as the layer becomes thinner, which is accompanied by progressively
poorer coupling between the wave and the metal; it is this poor coupling
that manifests itself in $\xi\to0$. The same poor coupling will thus
be characteristic for VPPs resulting from LRSPPs, which will therefore
be very similar in properties to plane waves propagating in the dielectric
of the HMM and thus occupy but a very narrow range of $\kappa$.

On the other hand, for $\xi$ to be on the order of unity in the LRSPP
case, the quantity $2\pi d_{m}/\lambda$ needs to be between 0.25
and 1 (see Fig.~\ref{FIG:Metal}d), i.e., the multilayer should not
be very subwavelength. In this regime, it can be expected that both
LRSPPs and SRSPPs may contribute to the VPP band, in line with the
recent observation that real multilayer structures can outperform
ideal HMMs for some values of layer thicknesses \cite{ourPRA12,ourOL11}.

\subsection{Graphene-dielectric multilayers}

Besides metallic layers, plasmonic excitations are present in other
thin-film structures such as monolayered graphene, and it has been
proposed that separating graphene layers by dielectric spacers and
combining them into multilayers \cite{grapheneMultXu} can give rise
to a new type of HMMs, predominantly in the THz range \cite{grapheneAndrei,grapheneBelov,grapheneStrangi,grapheneCapolino}.
Here we apply our approach to analyze the prerequisites needed for
a VPP band formation.

\begin{figure}[b]
\includegraphics[width=1\columnwidth]{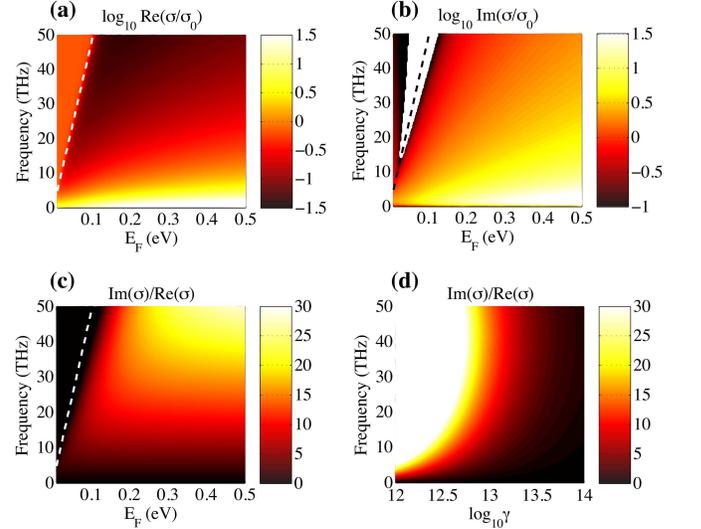}

\caption{(Color online) Graphene conductivity $\sigma$ in the units of elementary
conductivity $\sigma_{0}=e^{2}/4\hbar=0.061$mS in logarithmic scale,
(a) real and (b) imaginary part, depending on frequency $\omega$
and electrochemical potential (Fermi level) $E_{F}$. The dashed line
corresponds to the Pauli blocking limit $\hbar\omega=2E_{F}$. The
white region around the dashed line in (b) corresponds to the region
of negative $\mathrm{Im}\,\sigma$. Also shown is the figure of merit
$FoM=\mathrm{Im}\,\sigma/\mathrm{Re\,\sigma}$ dependence on frequency
and (c) $E_{F}$ for the damping $\gamma=10^{13}\mathrm{s^{-1}}$;
(d) on the damping $\gamma$ for the fixed Fermi level $E_{F}=0.2\mathrm{eV}$.
We consider the regions with $FoM>10$ suitable for HMMs.\label{FIG:GrapheneConductivity}}
\end{figure}

Graphene can be regarded as an infinitely thin sheet with surface
conductivity $\sigma$. In the THz to far infrared (far-IR) range
for graphene Fermi energy $E_{F}>k_{B}T$, it can be calculated according
to the Kubo approach with the formula \cite{HansonJAP}

\begin{equation}
\begin{gathered}\sigma=i\frac{e^{2}k_{B}T}{\pi\hbar^{2}}\left[\frac{E_{F}}{k_{B}T}+2\ln\left(1+e^{-\frac{E_{F}}{k_{B}T}}\right)\right]\frac{1}{\omega+i\gamma}\\
+i\frac{e^{2}}{4\pi\hbar}\ln\frac{2|E_{F}|-\hbar(\omega+i\gamma)}{2|E_{F}|+\hbar(\omega+i\gamma)}\end{gathered}
\label{eq:GrapheneCond}\end{equation}
where $T$ is the temperature and $\gamma$ is the damping rate, which
depends on the quality of graphene. The first and the second terms
is Eq.\,\eqref{eq:GrapheneCond} corresponds to interband and intraband
contributions, respectively. The resulting conductivity (real and
imaginary part) is shown in Fig.~\eqref{FIG:GrapheneConductivity}(a)--(b).
We see that the imaginary part $\mathrm{Im\,\sigma>0}$, which corresponds
to {}``metal-like'' behavior of graphene, everywhere except in a
narrow region of frequencies and electrochemical potential {[}white
area close to the dashed line in Fig.\,\ref{FIG:GrapheneConductivity}
(b){]}. As with the metal-dielectric multilayers, we will neglect
losses here, considering the frequency range where the real part of
$\sigma$ is small, and will assume that $\sigma$ is purely imaginary.

\begin{figure}
\includegraphics[width=1\columnwidth]{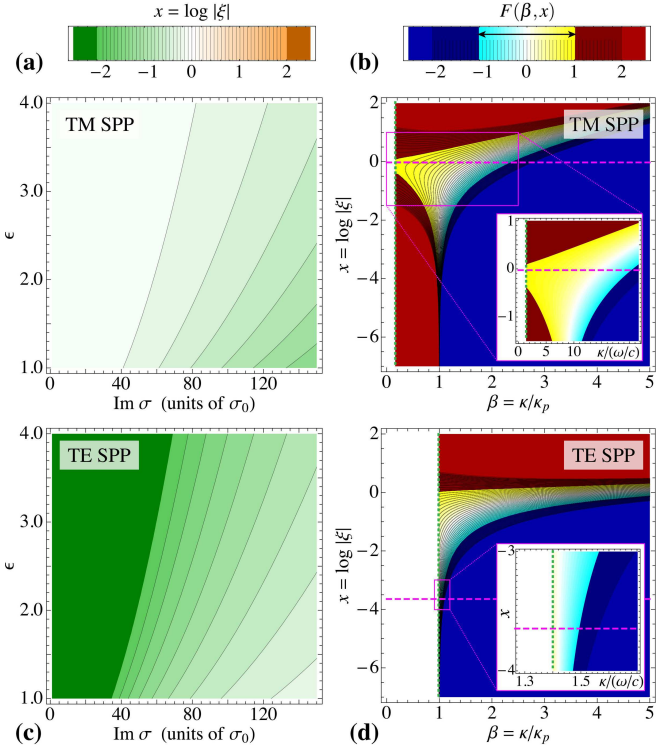}

\caption{(Color online) Formation of the VPP band in graphene HMMs. (a) The
dependence of $x$ on graphene conductivity and dielectric layer permittivity.
(b) Same as Fig.~\ref{FIG:Metal}b for graphene multilayers with
$\sigma=20i\sigma_{0}$ {[}where $\sigma_{0}=e^{2}/(4\hbar)${]},
$d=5$~nm, and $\epsilon=1.96$. (c--d) Same as (a--b) but for TE-polarized
graphene plasmons. \label{FIG:Graphene}}
\end{figure}

Surrounding a sheet of graphene with conductivity $\sigma$ by dielectric
with permittivity $\epsilon$, the transmission coefficient for the
TM polarization is given by \cite{graph-AndrOE}\begin{equation}
T=\frac{2\epsilon/\sqrt{\epsilon-\kappa^{2}}}{2\epsilon/\sqrt{\epsilon-\kappa^{2}}+(Z_{0}\sigma)},\label{eq:gr_RT}\end{equation}
where $Z_{0}=1/(\epsilon_{0}c)\approx377\,\Omega$ is the impedance
of free space. Assuming for now that there are no losses in graphene
and introducing $S\equiv2\epsilon/(Z_{0}\mathrm{Im\,}\sigma)$, Eq.~\eqref{eq:gr_RT}
yields the expressions for the pole expansion coefficients \begin{equation}
\kappa_{p}=\sqrt{\epsilon+S^{2}},\;\tau=-\frac{S^{2}}{\sqrt{\epsilon+S^{2}}},\;\xi=-\frac{S^{2}}{\epsilon+S^{2}}.\label{eq:gr_coefs}\end{equation}
It is remarkable that realistic graphene conductivities in the range
$\mathrm{Im\,}\sigma<100e^{2}/(4\hbar)$ yield large values of $S\simeq1\ldots100$.
Since $|\xi|$ tends to unity for large $S$, it turns out to be between
0.1 and 1, which is favorable for the VPP band to be broad and pronounced,
in a large parameter window, as shown in Fig.~\ref{FIG:Graphene}(a).
Changing the Fermi level allow the tuning of conductivity to the desired
value {[}see Fig.~\ref{FIG:GrapheneConductivity}(b){]} while maintaining
the deeply subwavelength thickness of such layers for THz frequencies;
Figure~\ref{FIG:Graphene}(b) additionally confirms the presence
of a VPP band in such graphene-dielectric multilayers. %
{} 

The key difference between VPPs in metal-dielectric and graphene multilayers
is that in the former $\xi\gtrsim1$ while in the latter $\xi\lesssim1$.
As a result, metal-dielectric layers benefit from a decrease of $\eta\kappa_{p}$
(e.g., by decreasing $d_{d}$ and making the structure more subwavelength),
whereas for graphene multilayers this is less relevant because the
slanted branch of $F(\beta,x)\in[-1;1]$ is outside of the working
values of $x<0$.

It was also pointed out recently that in addition to conventional
SPPs for TM-polarized waves, graphene supports transverse TE-polarized
SPPs \cite{TEgraphene2007,TEgrapheneOE} in a narrow range of parameters
where $\mathrm{\mathrm{Im\,}\sigma}<0$ and $\mathrm{Re\,}\sigma$
is small. It is interesting to analyze whether these SPPs can give
rise to HMM-like behavior. The TE counterparts to Eqs.~\eqref{eq:gr_RT}--\eqref{eq:gr_coefs}
are \begin{equation}
T'=\frac{2\sqrt{\epsilon-\kappa^{2}}}{2\sqrt{\epsilon-\kappa^{2}}+(Z_{0}\sigma)},\label{eq:gr_RT-S}\end{equation}
and (introducing $Q\equiv Z_{0}|\mathrm{Im\,}\sigma|/2$)\begin{equation}
\kappa'_{p}=\sqrt{\epsilon+Q^{2}},\;\tau'=\frac{Q^{2}}{\sqrt{\epsilon+Q^{2}}},\;\xi'=\frac{Q^{2}}{\epsilon+Q^{2}}.\label{eq:gr_coefs-S}\end{equation}
Here we note that TE-polarized plasmons only exist very close to the
singularity point in the graphene conductivity, with realistic $|\mathrm{Im\,}\sigma|<1\ldots2e^{2}/(4\hbar)$.
Hence $Q$ is a small quantity on the order of 0.05, making $\kappa'_{p}$
very close to $\sqrt{\epsilon}$ and $\xi'\ll1$. This makes TE-polarized
plasmons in graphene much like LRSPPs, which only hybridize into an
extremely narrow VPP band, as can indeed be seen in Fig.~\ref{FIG:Graphene}(c--d).

\subsection{Influence of losses}

It is important to note that in presence of losses, bulk plasmonic
waves in the high-$k$ band would acquire an imaginary part, and the
difference between propagating waves within the VPP band and evanescent
waves outside it becomes less pronounced. For smaller amounts of losses,
such as for $\mathrm{Im}\,\epsilon_{m}\ll|\mathrm{Re}\,\epsilon_{m}|$
in better metals, the high-$k$ band can nonetheless be rather well-defined,
even though its edges are smeared \cite{ourPRA12}, so the conclusions
of the present analysis would persist. 

However, in graphene there can be conditions when the real part of
the conductivity is quite significant {[}see Fig.\ref{FIG:GrapheneConductivity}(a){]}.
In such cases, the presented results can only be used as a guideline,
and the precise characteristics of VPPs should be established by additional
calculations.

To define the parameter range where the present analysis is applicable,
we introduce the figure-of-merit $FoM=\mathrm{Im}\,\sigma/\mathrm{Re}\,\sigma$,
which is an adaptation of the quantity commonly used to characterize
the amount of losses in metamaterials to a single graphene sheet .
We will assume that losses in graphene are small if $FoM>10$. Figure
\ref{FIG:GrapheneConductivity}(c) presents the figure-of-merit for
the damping value $\gamma=10^{13}\,\mathrm{s^{-1}}$ \cite{Tassin2013}.
%
{} As we see, $FoM>10$ corresponds to the Fermi level \emph{$E_{F}>0.15\mathrm{eV}$}
and frequencies above 20 THz, whereas in the lower THz and microwave
range graphene is essentially just a dissipative layer (resistor).
We should keep in mind that at photon energies larger than 0.2 eV,
which corresponds to the frequencies above 50 THz, the interaction
with the lattice phonons of the dielectric spacer layers in multilayered
graphene introduce additional large losses not taken into account
in Eq. \eqref{eq:GrapheneCond}. Therefore the region from 20 to 50
THz is probably the best for the realization of graphene based HMMs,
and larger $E_{F}$ are favorable for better HMM performance.

Another parameter, namely, damping (or collision frequency) $\gamma$
depends very much on the quality of graphene (its growth process and
handling when transferring to the substrate). The values reported
in the literature vary from $10^{12}\,\mathrm{s^{-1}}$ to $10^{14}\,\mathrm{s^{-1}}$
(the reader is referred to the recent review of graphene for THz applications\cite{Tassin2013}).
In Fig.~\ref{FIG:GrapheneConductivity}(d) the influence of damping
on the figure-of-merit is demonstrated. Whereas for the above mentioned
$\gamma=10^{13}\mathrm{s^{-1}}$ graphene could be only used for HMM
starting from 20 THz, reducing the damping by 10 times ($\gamma=10^{12}\,\mathrm{s^{-1}}$)
makes graphene HMMs feasible starting from as low as 1.6 THz. On the
other hand, doubling the damping to $\gamma=2\times10^{13}\,\mathrm{s^{-1}}$
makes graphene useless for building HMMs in the entire THz-IR range.
However, there are definite grounds for optimism in the constant progress
in graphene fabrication technology. For example, chemical vapor deposition
growth of centimeter-large monocrystalline graphene with the quality
rivaling that of exfoliated graphene \cite{grapheneFabScience} and
large mobility of carriers in graphene surrounded by two-dimensional
boron nitride \cite{grapheneFabBN} have been reported recently.

\section{Conclusions and outlook\label{sec:CONCLUSIONS}}

In summary, we have investigated the general theoretical conditions
for an arbitrary elementary excitation existing in the unit cell of
a multilayer periodic system to hybridize into a broadband bulk high-$k$
propagating waves (such as VPPs in HMMs). By isolating the unit-cell
elementary excitation in the form of a generalized resonance defined
by a pole-like response in its Fresnel reflection and transmission
coefficients {[}Eq.~\ref{eq:pole}{]}, and by using Bloch's theorem
to couple the unit cells via dielectric spacer layers, we have derived
analytic relations connecting the width of the resulting band of propagating
waves in the $k$-space with the properties of the elementary excitations,
such as the pole location and strength, as well as parameters of the
dielectric spacer layers. 

Using these analytical expressions, we have confirmed that one kind
of surface plasmons existing in thin metal layers, namely the SRSPPs,
can and normally do give rise to a broad band of volume plasmon polaritons,
resulting in HMM-like properties of subwavelength metal-dielectric
multilayers \cite{ourFocusOE13}. Conversely, the other kind of SPPs
in such layers, namely the LRSPPs, only form a very narrow plasmonic
band near the light line of the dielectric and do not produce a broad
high-$k$ band. 

We have also applied the formalism to multilayered graphene-dielectric
metamaterials in the THz range and shown that TM-polarized plasmons
in individual graphene sheets do hybridize to form VPPs with HMM-like
properties, and the VPP band is broadband enough for realistic values
of graphene conductivity {[}for the considered geometry $\mathrm{Im}\,\sigma<100e^{2}/(4\hbar)${]}.
On the other hand, transverse (TE-polarized) graphene plasmons only
form a very narrow VPP band, not giving rise to HMM properties and
behaving like LRSPPs in this respect. We have also shown that graphene
can be a good building material for high-$k$ band THz and IR metamaterials,
if it has sufficiently high quality (the damping $\gamma$ smaller
than $2\times10^{13}\mathrm{s^{-1}}$).

Along with providing the general theoretical understanding of the
formation of a high-$k$ band of bulk propagating waves from fixed-$k$
surface excitations in individual layers of a multilayer system, our
results have promising practical applications. They are twofold. First,
the analytic expressions allow for very easy and computationally efficient
estimations of VPP dispersion in existing metal-dielectric and graphene
multilayer HMMs, which can be used to design HMMs with optimized performance.
Second, on a more abstract level, the formalism provides insight into
a general question whether broadband large-wavevector higher-dimensional
response should be expected from \emph{any} given type of lower-dimensional
elementary excitations in \emph{arbitrary} periodic systems, not necessarily
bi-layer unit cells, but also many-layer and gradient. Examples may
include new types of photonic structures such as waveguide arrays
and multilayers based on Bloch surface waves or spoof surface plasmons.
Moreover, by virtue of mathematical similarities between electromagnetic
waves and other wave phenomena in physics (such as acoustic waves
in elastic multilayers and steady-state solutions of the Schr\"{o}dinger
equation in multiple quantum-well heterostructures), it can be speculated
that the present results may be applied to these alternative systems,
extending the metamaterial approach beyond electromagnetism.
\begin{acknowledgments}
S.V.Z. acknowledges financial support from the People Programme (Marie
Curie Actions) of the European Union's 7th Framework Programme FP7-PEOPLE-2011-IIF
under REA grant agreement No.~302009 (Project HyPHONE). A.A. acknowledges
financial support from the Danish Council for Technical and Production
Sciences through the GraTer (0602-02135B) project. J.E. Sipe acknowledges
financial support from the Natural Sciences and Engineering Research
Council of Canada.\end{acknowledgments}

\end{document}